\documentclass[reprint,twocolumn,showpacs,amsmath,amssymb,superscriptaddress,aps,prl]{revtex4-1}

\usepackage{graphicx}
\usepackage[english]{babel}
\usepackage{upgreek}	
\usepackage{etoolbox} 

\newcommand{\var}{\ensuremath{\text{var}}}

\newcommand{\ket}[1]{\ensuremath{|#1\rangle}}
\newcommand{\mean}[1]{\ensuremath{\langle#1\rangle}}
\newcommand{\std}[2][]{\ifstrempty{#1}{\sigma_{#2}}{\sigma_{#2,#1}}}

\begin{document}

\title{Quantum metrology with a scanning probe atom interferometer}

\author{Caspar F. Ockeloen}
\author{Roman Schmied}
\author{Max F. Riedel}
\author{Philipp Treutlein}\email{philipp.treutlein@unibas.ch}
\affiliation{Department of Physics, University of Basel, \\Klingelbergstrasse 82, 4056 Basel, Switzerland}

\begin{abstract}
We use a small atomic Bose-Einstein condensate as an interferometric scanning probe to map out a microwave field near a chip surface with a few micrometers resolution. Using entanglement between the atoms we overcome the standard quantum limit of interferometry by 4~dB and maintain enhanced performance for interrogation times up to 20~ms. This demonstrates the usefulness of quantum metrology with entangled states when the particle number is limited due to the small probe size. Extending atom interferometry to micrometer spatial resolution enables new applications in electromagnetic field sensing, surface science, and the search for fundamental short-range interactions.
\end{abstract}

\maketitle

Interferometers operating with large atomic ensembles offer unsurpassed precision in measurements of inertial forces, atomic properties, and fundamental constants, and currently define the standard of time \cite{Cronin2009,Kitching2011}. 
Using a small atomic cloud as a scanning probe interferometer with high spatial resolution would enable new 
applications in electromagnetic field sensing, surface science, and the search for fundamental short-range interactions \cite{Cronin2009}.
However, as a small probe necessarily contains only a small number of atoms, the standard quantum limit (SQL) due to projection noise \cite{Itano1993} places a tight bound on the achievable precision.
We report a scanning probe atom interferometer which overcomes the SQL by entangling the atoms, and use it for a high-resolution microwave field measurement. 
With a spin-squeezed Bose-Einstein condensate \cite{Riedel2010,Gross2010} (BEC) on an atom chip \cite{ReichelVuletic2011} we achieve a precision of $4\,\text{dB}$ in variance below the SQL. Performance below the SQL is maintained for interrogation times up to $20\,\text{ms}$, more than an order of magnitude longer than in previous experiments \cite{Gross2010,Louchet-Chauvet2010,Leroux2010}. We transport our $1.1\,\upmu\text{m}$-sized probe between $40\,\upmu\text{m}$ and $16\,\upmu\text{m}$ from the atom chip surface, and record the spatial profile of a microwave magnetic field with sub-SQL precision. 
Our experiment demonstrates the usefulness of quantum metrology with entangled states \cite{Giovannetti2004} in a situation where a physical process strictly constrains the number of available particles.
The technique developed here is relevant for high-resolution imaging of electromagnetic fields near micro-fabricated structures \cite{Lin2004,Obrecht2007,Aigner2008,Bohi2010,Gierling2011}.

In an atom interferometer, the external (motional) or internal (spin) state of atoms is coherently split and allowed to follow two different pathways \cite{Cronin2009,Kitching2011}. During the interrogation time, a phase $\varphi$ between the paths is accumulated, which depends on the quantity to be measured. When the paths are recombined, the wave-character of the atoms gives rise to an interference pattern, from which $\varphi$ can be determined. In recording this interference, however, the particle-character of the atoms is revealed, as a measurement randomly projects the wave function of each atom into a definite state. When operating with an ensemble of $N$ uncorrelated (non-entangled) atoms, the resulting binomial statistics of counting individual atoms in the output states limits the phase uncertainty of the interferometer to $\std\varphi \geq 1/\sqrt{N}$, the standard quantum limit (SQL) of interferometric measurement \cite{Itano1993}.

\begin{figure}%
\includegraphics[width=\columnwidth]{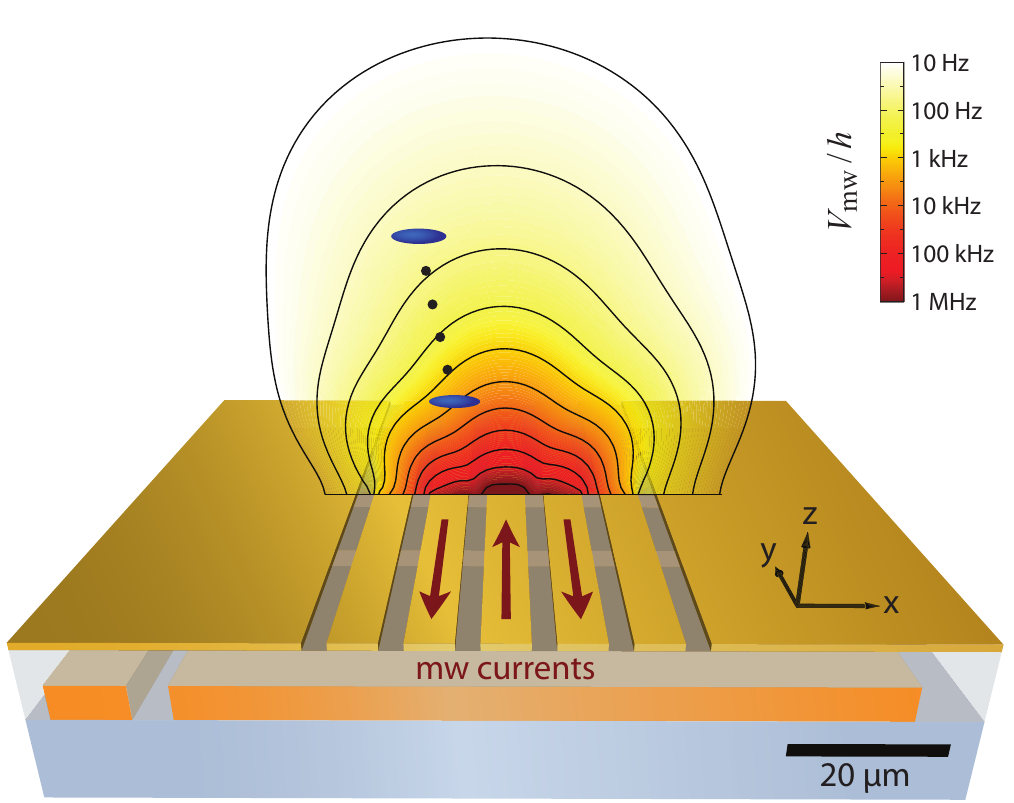}
\caption{{Experimental setup.} Central region of the atom chip showing the atomic probe (blue, size to scale) and the scanning trajectory we use (between $40\,\upmu\text{m}$ and $16\,\upmu\text{m}$ from the chip surface).
The probe is used to measure the magnetic near-field potential generated by an on-chip microwave guide (microwave currents indicated by arrows). A simulation of the potential is shown in red/yellow.}
\label{fig:setup}%
\end{figure}

An obvious way to reduce $\std\varphi$ is to increase the number of particles in the interferometer. Most atom interferometers thus operate with large ensembles containing millions of atoms. However, there are important cases where a physical process places an upper bound on $N$, so that this strategy is not feasible.
In chip-based atomic clocks, for example, undesired collisional frequency shifts limit the usable atomic density and thus the atom number to $N\lesssim 10^5$ \cite{Rosenbusch2009}. A particularly tight limit applies to atom interferometry with high spatial resolution, where the required small probe size gives rise to an upper bound on $N$ due to density-dependent collisional trap losses. As an example, three-body recombination in a $^{87}$Rb BEC places a limit of $N\leq 400$ on the atom number in a probe volume of  $1\,\upmu\text{m}^3$ if a trap lifetime $\geq 1\,\text{s}$ is desired \cite{Burt1997}.

Quantum metrology provides a way to overcome the SQL at fixed $N$ by using entanglement between the probe particles \cite{Giovannetti2004}. This allows reducing the phase measurement uncertainty towards the ultimate Heisenberg limit, $\std\varphi \geq 1/N$. An important class of interferometrically useful entanglement is provided by spin-squeezed states \cite{Kitagawa1993,Sorensen2001a}, characterized by the squeezing parameter \cite{Wineland1994} $\xi<1$, defined such that $\std\varphi = \xi/\sqrt{N}$. Recent experiments with atomic ensembles have demonstrated suitable entangled states \cite{Esteve2008,Appel2009,Schleier-Smith2010,Riedel2010,Lucke2011,Gross2011,Sewell2011,Hamley2012}, and with optically trapped atoms full interferometer sequences have been implemented \cite{Gross2010,Louchet-Chauvet2010,Leroux2010}. For scanning probe measurements near a surface, magnetic traps on atom chips have proven to be useful, as they provide sub-micrometer position control over small BECs \cite{Lin2004,Obrecht2007,Aigner2008,Hunger2010,Gierling2011}. However, quantum metrology with entangled states has not yet been demonstrated in such experiments.

Our interferometric scanning probe is a $^{87}$Rb BEC prepared close to the surface of an atom chip. The two ground-state hyperfine levels $\ket{1} = \ket{F=1, m_F=-1}$ and $\ket{2} = \ket{F=2, m_F=1}$ serve as interferometer pathways. 
Figure~\ref{fig:setup} shows an overview of our experimental setup \cite{Riedel2010,methods}. 
The atom chip has two layers of wires carrying both continuous (dc) and microwave (mw) currents. 
The dc currents generate magnetic traps, which are identical for both states.
In each run of the experiment we start by preparing a pure BEC of $N = 1400\pm40$ atoms (shot-to-shot preparation noise) in state $\ket{1}$, in a  trap situated $40\,\upmu\text{m}$ away from the chip surface. 
This trap can be smoothly shifted towards the chip surface to perform high-resolution measurements. The probe size is characterized by the BEC radii of $R_y = R_z = 1.1\,\upmu\text{m}$ and $R_x = 4.0\,\upmu\text{m}$.

To create entanglement between the atoms, we make use of two-body collisions. Normally, these collisions have negligible effect on the internal state of the BEC, due to a coincidence of the scattering lengths of our state pair. A crucial feature of our experiment is that we can ``turn on'' the effect of collisions for a well-defined time by spatially separating the states \cite{Li2009}. For this we use a state-dependent potential, which is generated by the on-chip microwave currents. By recombining the states, the collisions are effectively ``turned off'' again, so that they do not perturb the subsequent interferometric sequence. In Ref.~\onlinecite{Riedel2010} we have used this technique to produce spin-squeezed states of a BEC.

\begin{figure}%
\includegraphics[width=\columnwidth]{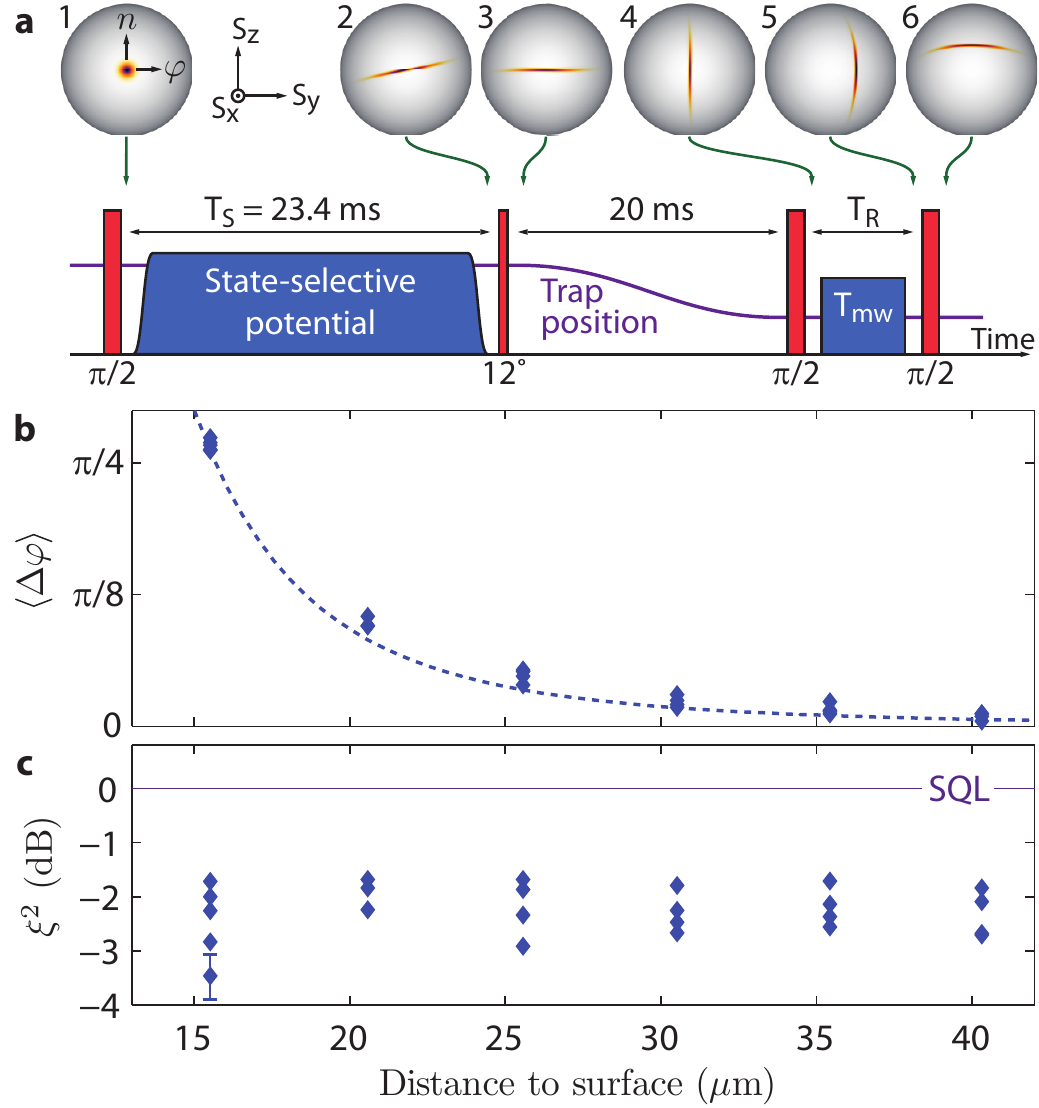}
\caption{{Scanning probe interferometer operating below the SQL.} (a) Experimental sequence, showing Rabi (red) and on-chip microwave (blue) pulses and the trap position (purple). Spheres 1-6 show the Wigner function of the collective spin state at various stages of the experiment, simulated for $N=200$ atoms. (b) Measured phase shift $\mean{\Delta\varphi}$ induced by a microwave near-field pulse as a function of the atom--surface distance, compared to the simulated potential (dotted line, see also Fig.~\ref{fig:setup}). (c) Measured performance of the interferometer, expressed as squeezing factor $\xi^2$. Each data point (based on 240 measurements) has a statistical uncertainty of $\pm0.4\,\text{dB}$, shown as error bar on the lower left point. The experiment was repeated up to 5 times at each position.}
\label{fig:scanningprobe}%
\end{figure}

The internal state of our BEC can be described by a collective spin $\vec{S}$, whose quantum state is visualized on a sphere (Fig.~\ref{fig:scanningprobe}a). The $S_z$-component is proportional to the population difference $n=(N_2-N_1)/(N_1+N_2)$, where $N_i$ is the atom number in state $\ket{i}$. The azimuthal angle is the relative phase $\varphi$ between $\ket{1}$ and $\ket{2}$.
Using mw and radio-frequency (rf) Rabi pulses generated off-chip, we can coherently couple $\ket{1}$ and $\ket{2}$, thereby rotating the state on the sphere. The effect of collisions is well described by the Hamiltonian $H_\text{int} = \hbar\chi S_z^2$, which creates spin-squeezing by ``twisting'' the state on the sphere \cite{Kitagawa1993,Sorensen2001a,Li2009}. The rate $\chi$ is controlled by the state-dependent potential as described above.

The experimental sequence of our scanning probe interferometer is shown in figure~\ref{fig:scanningprobe}a. 
We start with a $\pi/2$-pulse, creating a coherent spin state (sphere 1) on the equator of the sphere. Then, we apply $H_\text{int}$ for $T_\text{S} = 23.4\,\text{ms}$ by turning on the state-dependent potential. The result is a spin-squeezed state (sphere 2) with a measured $\xi^2 = -4.3\pm0.4\,\text{dB (statistical)}\pm0.4\,\text{dB (systematic)}$. The systematic uncertainty arises from calibrating the imaging system and applies to all measurements of $\xi^2$ in this paper \cite{methods}. 
Next, we apply a $-12^\circ$ rotation around the $S_x$ axis, aligning the anti-squeezed quadrature with the equator (sphere 3). This renders the state minimally sensitive to phase noise during the following $20\,\text{ms}$, in which we transport the atoms to the position where the interferometric measurement is to be made \cite{methods}. Finally, we perform a full Ramsey interferometer sequence consisting of a $\pi/2$-pulse around the state's center to make it maximally phase sensitive (sphere 4), an evolution time $T_{\text{R}}$ during which the phase $\varphi$ is accumulated (sphere 5), and a final $\pi/2$-pulse mapping $\varphi$ onto $n$ (sphere 6). 
We read out the population difference, which oscillates as $n = C\sin\varphi$, where $C\le1$ is the interferometric contrast. Experimentally, an offset phase $\theta$ can be added to the last $\pi/2$-pulse. For example, we can choose $\theta$ such that $\mean{n}=0$ and the interferometric phase uncertainty is $\std{\varphi} = \std{n}/C$, where $\sigma$ denotes standard deviation.

We demonstrate the scanning probe interferometer with spatially resolved measurements of the on-chip microwave near-field at $6.8$\,GHz. 
We use $T_\text{R} = 100\,\upmu\text{s}$, during which we pulse on the field for $T_{\text{mw}} = 80\,\upmu\text{s}$ with a detuning of $12\,\text{MHz}$ above the $\ket{F=1,m_F=0} \rightarrow \ket{F=2,m_F=0}$ transition. 
This results in an additional phase shift $\Delta\varphi = T_{\text{mw}} V_{\text{mw}}/\hbar$, where $V_{\text{mw}}$ is the differential ac Zeeman shift of states $\ket{1}$ and $\ket{2}$ \cite{methods}. 
Part b of figure \ref{fig:scanningprobe} shows such measurements at the positions indicated in figure~\ref{fig:setup}, between $40\,\upmu\text{m}$ and $16\,\upmu\text{m}$ from the chip surface. At each position we measure $\varphi$, and the reference phase $\varphi_\text{0}$ in a separate measurement without mw pulse. The mean phase shift due to the mw is given by $\mean{\Delta\varphi} = \mean{\varphi} - \mean{\varphi_{\text{0}}}$. We extract these mean phases from fits of sine functions to the measured values of $n$ as a function of $\theta$. 
The measured shape of $V_{\text{mw}}$ agrees well with a simulation based on previous measurements with a different technique \cite{Bohi2010}.
Figure~\ref{fig:scanningprobe}c shows the performance of our interferometer in terms of the squeezing factor $\xi^2$, measured while operating at $\mean{n} = 0$ with mw pulse on. For all positions, our interferometer performs well below the SQL, with an average performance of $\mean{\xi^2} = -2.2\,\text{dB}$ corresponding to a single-shot phase sensitivity of $\std{\varphi} = 1.2^\circ$, or a microwave field sensitivity of $\delta B = 2.4~\upmu$T.

In general, interferometric sensitivity scales linearly with $T_\text{R}$, making long interrogation times advantageous for high precision. 
We therefore measure the performance of our interferometer as a function of $T_\text{R}$, similar to \cite{Leroux2010}.
We use the same sequence as described above, but omit transporting the atoms. Instead, the second and third Rabi pulses are combined into a single $+78^\circ$ rotation, which aligns the squeezed quadrature with the equator and immediately starts the free evolution. During $T_\text{R}$, no additional phase shift is applied.

\begin{figure}
\includegraphics[width=\columnwidth]{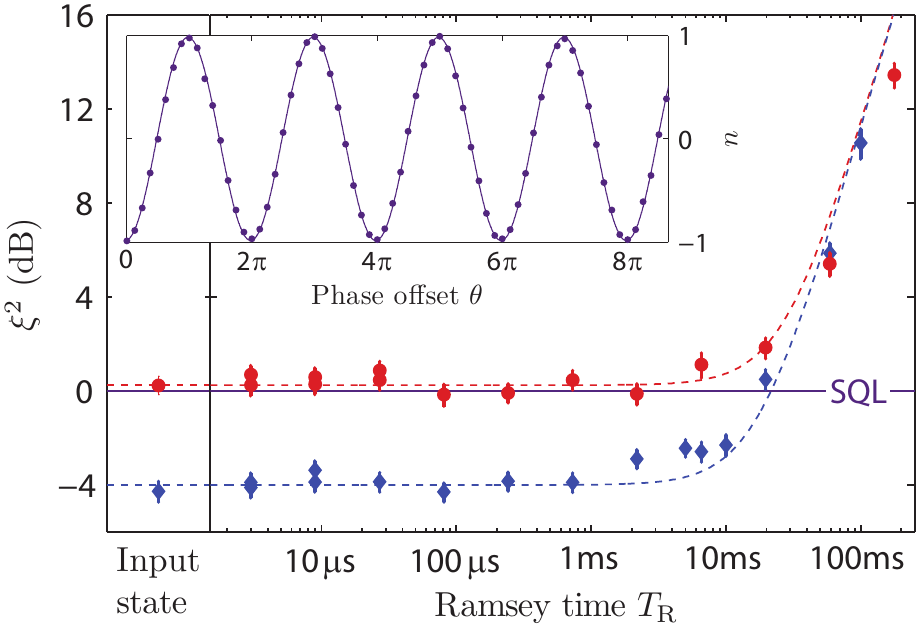}
\caption{{Interferometer performance.} Observed phase noise in a Ramsey interferometer with squeezed (blue circles) and coherent (red diamonds) input states for varying interrogation times $T_\text{R}$. Dashed lines model constant performance of $\xi^2 = -4\,\text{dB}$ (squeezed state) and $+0.2\,\text{dB}$ (coherent state), plus technical noise due to shot-to-shot frequency fluctuations. Each data point is the result of 240 measurements, and error bars indicate statistical uncertainty. Inset: typical squeezed-state Ramsey fringe measurements for $T_\text{R} = 5\,\text{ms}$ (points) and fitted sine (line) yielding a contrast of $C = (98.1\pm0.2)\%$.}
\label{fig:interferometer}%
\end{figure}

Figure \ref{fig:interferometer} shows the measured phase uncertainty for varying $T_{\text{R}}$. The interferometer operates at $\xi^2 \approx -4\,\text{dB}$ below the SQL, maintaining the squeezing level of the input state. It performs better than the SQL up to $T_{\text{R}} = 20\,\text{ms}$, an improvement by a factor of 30 compared to previous results \cite{Leroux2010}.
In a reference measurement using a coherent state, our performance is consistent with the SQL plus independently measured detection noise of $\std[\text{det}]{n} = 5.1\times10^{-3}$ ($5$ times smaller than the SQL). After $T_{\text{R}} > 20\,\text{ms}$, both measurements are limited by technical noise, consistent with shot-to-shot fluctuations of $150\,\text{mHz}$ (rms) in the relative frequency between our reference oscillator and the atomic resonance. 

After an interrogation time of $20\,\text{ms}$, our measured noise level corresponds to a single-shot sensitivity in the ac Zeeman shift of $\delta V_\text{mw}/h=0.21\,\text{Hz}$. We now estimate the corresponding microwave field sensitivity for a near-resonant field, obtaining a noise-equivalent field amplitude of $\delta B \approx \delta V_\text{mw} / \mu_\text{B}=15$\,pT. Taking our experimental cycle of $11\,\text{s}$ into account, we obtain a microwave field sensitivity of $50\,\text{pT}/\sqrt{\text{Hz}}$. We emphasize that this sensitivity is achieved with a small probe of $20\,\upmu\text{m}^3$ in volume. We have recently developed non-interferometric microwave field imaging techniques with ultracold \cite{Bohi2010} and room-temperature \cite{Bohi2012} atomic vapors. Compared to these works, the entanglement-enhanced interferometric sensor demonstrated here is about three orders of magnitude more sensitive and offers higher spatial resolution.

In conclusion, we have experimentally demonstrated a scanning probe atom interferometer operating beyond the standard quantum limit, and used it for the measurement of a microwave near-field. This is the first demonstration of entanglement-enhanced atom interferometry with a high spatial resolution scanning probe, and promises further high-resolution sensing and measurement applications. 
A particularly important challenge is to demonstrate similar scanning capability and sub-SQL operation with an atom interferometer that measures inertial forces or surface potentials on a small length scale \cite{Cronin2009}. This requires that the atomic wave functions are spatially split during the interrogation time $T_\text{R}$. Our atom chip allows for the implementation of such a sequence \cite{Bohi2009}, but an important additional requirement for sub-SQL operation is to suppress the nonlinearity $\chi$ in the split state while the interferometer phase is accumulated. 
One strategy to achieve this is to decrease the trap frequency after splitting the trap, as $\chi$ scales with the inverse density. An alternative is to choose a different atomic species where a Feshbach resonance can be used to tune the interaction strength to zero \cite{Gustavsson2008,Fattori2008}.

\begin{acknowledgments}
We thank P.~B{\"o}hi for help with the experimental setup. This work was supported by the Swiss National Science Foundation and by the EU projects AQUTE and QIBEC.
\end{acknowledgments}



\section*{Supplemental Material}

Our experimental setup has been previously described \cite{Riedel2010,Bohi2010,Bohi2009,RiedelThesis}.
We use a ``dimple'' trap, where transverse confinement comes from a chip wire carrying $I_\text{L} = 130\,\text{mA}$ in $x$-direction plus a static field $B_y = 5.2\,\text{G}$, and longitudinal confinement from three ``dimple'' wires carrying in total $I_\text{D} = -2\,\text{mA}$ in the $y$-direction. The dimple currents are asymmetrically distributed to set the trap position at $x_0 = -13\,\upmu\text{m}$ from the wire center \cite{RiedelThesis}. States $\ket{1}$ and $\ket{2}$ see identical dc potentials, with trap frequencies $f_x = 115\,\text{Hz}$ and $f_y = f_z = 540\,\text{Hz}$. We create a cold cloud with stable atom number by evaporative cooling to far below the BEC transition.

To shift the trap, we define a parameter $\eta$, which we vary from $\eta=1$ (original trap) to $\eta = 0.5$, and scale $I_\text{L}\propto\eta^2$, $I_\text{D}\propto\eta^4$ and $B_y\propto\eta$, while keeping $B_x=3.2\,\text{G}$ constant. This keeps the trap geometry constant within $10\%$, and the magnetic field at the trap bottom within $50\,\text{mG}$ of the ``magic'' value of $3.23\,\text{G}$, where the differential Zeeman shift between $\ket{1}$ and $\ket{2}$ vanishes to first order \cite{Treutlein2004}. We transport the atoms by smoothly ramping all parameters from $\eta=1$ to their final value during $20\,\text{ms}$. We simulate the magnetic trap geometry with estimated position accuracy of $1\,\upmu\text{m}$ in $x$ and $y$-direction, and $2\,\upmu\text{m}$ in $z$-direction. We estimate the size of the BEC based on Ref.~\onlinecite{Mateo2007}.

We drive coherent Rabi rotations between states $\ket{1}$ and $\ket{2}$ using a two-photon mw + rf transition, with both fields generated off-chip. We detune the mw by $500\,\text{kHz}$ above the $\ket{F=2,m_F=0}$ intermediate state. Our two-photon Rabi frequency varies from $550\,\text{Hz}$ at $\eta=1$ to $1.2\,\text{kHz}$ at $\eta = 0.5$, which we attribute to induced mw and rf currents in the chip structures. All signals are derived from an ultra-stable quartz oscillator (short term stability $\sigma_\tau < 10^{-12}/\sqrt\text{Hz}$), and we achieve a fidelity of $(99.74\pm0.04)\%$ for a single $\pi/2$-pulse.

We define a collective spin $\vec{S} = \sum_{i=1}^{N} \vec{s}_i$, where the pseudo-spin-1/2 $\vec{s}_i$ describes the internal state of atom $i$. The states can be represented by their spherical Wigner functions \cite{Schmied2011}. To visualize the effect of squeezing (Fig. 2a, sphere 2), we simulate evolution under $H_\text{int}$, starting from a coherent state with $N=200$ and $\mean{S_z} = 0$. The evolution time is chosen to match the angle of maximum squeezing with our experimental data.

The dimple wires double as waveguide for mw signals. We use far off-resonant mw, blue-detuned by $12\,\text{MHz}$ from the $\ket{F=1,m_F=0} \rightarrow \ket{F=2,m_F=0}$ transition, creating a differential level shift between $\ket{1}$ and $\ket{2}$ of $(71,46,39)\,\text{kHz/G}^2$ for the $(\pi,\sigma^+,\sigma^-)$-components of the mw magnetic field amplitude. The strong spatial gradients of the mw field give rise to a state-dependent force.

During $T_\mathrm{S}$, we apply a mw current of $\approx 15\,\text{mA}$ (rms), separating the potential minima of $\ket{1}$ and $\ket{2}$ by $\approx 140\,\text{nm}$. After turning on the potential (with a $350\,\upmu\text{s}$ smooth ramp), the spatial wave functions oscillate in their new potentials. However, collisional interactions push the states further apart (coherent demixing) and slow down the oscillation. We turn off the mw field when the states maximally overlap after one oscillation.

During $T_\mathrm{mw}$, we apply a mw current of $\approx 5\,\text{mA}$ to generate $V_\text{mw}$. This short pulse does not significantly affect the spatial wave functions. In Figs.~1 and 2b we show a simulation of $V_\text{mw}$ based on previous measurements \cite{Bohi2010}. Since our present measurements are more precise, we use them to calibrate the efficiency of coupling the mw current onto the chip. The resulting global calibration factor agrees well with the measured mw transmission through the chip. The asymmetry in $V_\text{mw}$ arises from mw currents induced in chip wires next to the waveguide. 

The precise and well-calibrated determination of both $N_1$ and $N_2$ is essential to our measurements. We use state-selective absorption imaging \cite{Riedel2010}, after pushing the atoms away from the chip with a magnetic field gradient and a short time of flight (TOF, up to $7\,\text{ms}$). We experimentally optimize the exposure time ($70\,\upmu\text{s}$) and probe intensity ($1.3\,\text{mW/cm}^2$) for best signal-to-noise ratio. To reduce the effect of photon shot noise, we apply the algorithm described in Ref.~\onlinecite{Ockeloen2010}. We obtain $N_1$ and $N_2$ by integrating the atomic density profiles \cite{Reinaudi2007} over a region containing $>95\%$ of the atomic distribution. We achieve noise levels of $\std[\text{det}]{N_1} = 5.7$ and $\std[\text{det}]{N_2} = 4.2$ atoms, corresponding to $\std[\text{det}]{n} = (5.1\pm0.4)\times10^{-3}$. Trap positions close to the chip surface need longer TOF, resulting in $\std[\text{det}]{n} = 6.5\times10^{-3}$ for $\eta=0.5$.

We use the method of Ref.~\onlinecite{Reinaudi2007} to verify that our probe light is consistent with pure $\sigma^+$-radiation. We calibrate the atom number by measuring the shot noise of a coherent state (created by a single $\pi/2$-pulse) for varying $\mean{N}$, and compare to the model $\mean{N}^2\var(n) = \alpha\mean{N} + \std[\text{det}]{N_1}^2 + \std[\text{det}]{N_2}^2$. While for a coherent state we expect $\alpha=1$, a linear fit to the data (including the independently measured detection noise at $\mean{N}=0$) yields $\alpha=0.82\pm0.07$. We attribute this to an error in calculating the density profiles, as our imaging system only partially resolves the clouds, and correct for it by dividing $N_1$ and $N_2$ by $\alpha$.

We verify the linearity of the imaging by comparing a high-resolution Ramsey measurement ($T_\text{R} = 100\,\upmu\text{s}$) to the expected sinusoidal fit, and find that the slope of our Ramsey fringes deviates by at most $2\%$ from the model. Together with the calibration uncertainty, this results in a total systematic uncertainty of $10\%$ ($0.4\,\text{dB}$) in $\xi^2$.

Even though we achieve small shot-to-shot fluctuations in $N$, our interferometer is sensitive to a density-dependent level shift arising from mean-field interactions. We measure this shift to be $(5.1\pm0.1)\,\text{mHz/atom}$, based on the coherent-state reference data used in figure 2. We can measure $N$ much better than the preparation noise, and we correct for the shift for each shot individually \cite{Rosenbusch2009}. Without this correction, the scanning probe interferometer would still perform below the SQL at $\mean{\xi^2} = -1.7\,\text{dB}$, and our interferometer would remain sub-SQL for $10\,\text{ms}$. 


%
%
%
%
%
%
%
%
%
%
%

\end{document}